\documentclass[%
 reprint,
 amsmath,
 amssymb,
 pre
]{revtex4-1}
\usepackage{color}
\usepackage{multirow}
\usepackage{graphicx}
\usepackage{subfig}
\usepackage{dcolumn}
\usepackage{float}
\usepackage{bm}
\usepackage{makecell}

\begin{document}

\preprint{APS/123-QED}

\title{Wireless Communication Using Chaotic Baseband Waveform}

\author{Jun-Liang Yao}
\author{Yu-Zhe Sun}%
\author{Hai-Peng Ren}%
\altaffiliation{Corresponding author: renhaipeng@xaut.edu.cn}
\affiliation{Shaanxi Key Laboratory of Complex System Control and Intelligent Information Processing, Xi'an University of Technology, Xi'an 710048, China}
\author{Celso Grebogi}
\affiliation{Institute for Complex System and Mathematical Biology, University of Aberdeen AB24 3UE, United Kingdom}


\begin{abstract}
Some new properties of the chaotic signal have been implemented in communication system applications recently. They include: (i) chaos is proven to be the optimal communication waveform in the sense of very simple matched filter being used to achieve maximum signal to noise ratio; (ii) the amount of information contained in a chaotic signal is unaltered by a wireless multipath channel; and (iii) chaos property can be used to resist multipath effect. All these support the application of chaos in a practical communication system. However, due to the broadband property of the chaotic signal, it is very difficult for a practical transducer or antenna to convert such a broadband signal into a signal that would be suitable for practical band-limited wireless channel. Thus, the use of chaos property to improve the performance of conventional communication system without changing the system configuration becomes a critical issue in communication with chaos. In this paper, chaotic baseband waveform generated by a chaotic shaping filter is used to show that this difficulty can be overcome. The generated continuous-time chaotic waveform is proven to be topologically conjugate to a symbolic sequence, allowing the encoding of arbitrary information sequence into the chaotic waveform. A finite impulse response filter is used to replace the impulse control in order to encode information into the chaotic signal, simplifying the algorithm for high speed communication. A wireless communication system is being proposed using the chaotic signal as the baseband waveform, which is compatible with the general wireless communication platform. The matched filter and decoding method, using chaos properties, enhance the communication system performance. The Bit Error Rate (BER) and computational complexity performances of the proposed wireless communication system are analyzed and compared with the conventional wireless systems. The results show that the proposed chaotic baseband waveform of our wireless communication method has better BER performance in both the static and time-varying wireless channels. The experimental results, based on the commonly-used wireless open-access research platform, show that the BER of the proposed method is superior to the conventional method under a practical wireless multipath channel.
\end{abstract}

\pacs{Valid PACS appear here}
\maketitle
\section{Introduction}
As an important issue in nonlinear science, chaos has been found in many traditional fields, such as weather forecast, biomedicine, and control systems. Owing to some of its intrinsical properties, chaos has attracted lots of attentions also in the communication fields\cite{Williams2001Chaotic,Li2003A,ERIK2003REVIEW,Chen2017Design}. Since chaotic synchronization was proposed in \cite{Pecora1990Synchronization}, chaotic communication has been extensively investigated for more than twenty-five years.

One chaotic communication scheme is to use chaos as a carrier waveform or spread spectrum waveform in dealing with issues such as chaotic masking\cite{Cuomo2013Circuit}, Chaos Shift Keying (CSK)\cite{Dedieu2002Chaos,Yang2017Multi}, and Chaos-based Direct sequence Code Division Multiple Access (CD-CDMA)\cite{Kurian2005Performance}. In these applications, the chaotic waveform does not contain information, instead, the information is masked in the chaotic signal (for chaotic masking) or information is encoded using the key shift (for CSK). At the receiver, the complex chaotic synchronization is usually needed to recover the information. To avoid the difficulty of the precise synchronization, the Differential CSK (DCSK)\cite{Kolumb1996Differential,Fang2017A,Kaddoum2016Wireless} was proposed. The DCSK system is robust under practical wireless channel and exhibits lower implementation complexity\cite{Kaddoum2012Implementation}. However, the weaknesses of DCSK are high energy consumption, low data rate, weakened information security and requirement of widewidth delay lines. Subsequently, several variants of DCSK were developed to improve the performance of DCSK in different ways, as shown in Table I. However, these variants of DCSK have low bandwidth efficiency as compared to conventional wireless communication.
\begin{table}\label{VariantsOfDCSK}
 \centering
   \caption{Variants of DCSK}
   \begin{tabular}{|p{1.7cm}|p{6.2cm}|} 		        \hline
		Improvement& \multicolumn{1}{c|}{Methods}\\ \hline
		\multirow{5}{*}{\makecell[tl]{higher \\ data rate}} & Frequency-modulated DCSK
                                     \cite{KOLUMBAN1998FM,Kennedy2000Performance} \\
                                   & M-ary DCSK \cite{Wang2010MDCSK,Wang2015Design},\cite{Cai2016A,Cai2017Design} \\
                                   & High efficiency DCSK \cite{Yang2012High}\\
                                   & Multi-carrier DCSK \cite{Kaddoum2013Design}\\
                                   & Short reference DCSK \cite{Kaddoum2016Design2,Bai2018Chaos}\\  \hline
        \multirow{4}{*}{\makecell[tl]{lower bit \\error rate}} & Noise reduction DCSK \cite{Kaddoum2016NR}\\
                                           & Improved DCSK \cite{Kaddoum2015I}\\
                                           & Coded modulation DCSK \cite{Chen2018A} \\
                                           & DCSK based on matched filter \cite{Bai2018Experimental} \\
                                           \hline
        \multirow{2}{*}{\makecell[tl]{without \\ delay lines}} & Code-shift DCSK \cite{Xu2017Code}\\
                                     & Reference modulated DCSK \cite{Yang2013Reference}\\  \hline
        \multirow{2}{*}{\makecell[tl]{improved \\security}} & Permutation index DCSK \cite{Herceg2018Permutation,Ren2017A}\\
                                     &  \\ \hline
        \multirow{3}{*}{Others} & Continuous mobility DCSK \cite{Escribano2016Design}\\
        & DCSK-based automatic repeat request \cite{Fang2015Design}\\
        & Relay-based DCSK \cite{Fang2013Performance,Cai2018Design}\\ \hline

	\end{tabular}
 \end{table}

Another chaotic communication scheme is to encode the information bits in the chaotic waveform itself by controlling the dynamics of the chaotic system \cite{Hayes1993Communicating,Hayes1994Experimental,Kaddoum2012An,Kaddoum2013Spread}. At the receiver, the message can be decoded with a proper symbolic partition. This scheme has high power spectral density and does not need chaos synchronization \cite{Kaddoum2012An,Kaddoum2013Spread}. However, there are some encoding constraints using just the traditional chaotic dynamics\cite{ERIK2003REVIEW}, and it is difficult to encode arbitrary information bit sequence with low computational complexity. To overcome these difficulties, the noise filter was developed by using a chaos property \cite{Rosa1997Noise}, and the arbitrary information bits could be encoded by using a hybrid chaotic system and impulse control\cite{Ren2016Chaos, Ren2013}. Along this way, the intersymbol interference caused by multipath was decreased using a chaos property \cite{Yao2017PRE}.

In practice, chaos was reported to be successfully used in a commercial optical fiber communication system to get higher bit transmission rate \cite{Argyris2005Nature}. The chaotic pulse has also been considered as a possible solution for the low-rate and ultra wideband applications, which has been included in the wireless personal area network standard IEEE 802.15.4a\cite{Hyung2005IEEE802.15}. The frequency-modulated DCSK has been proposed as the wireless body area network standard IEEE 802.15.6\cite{Ieee2012IEEE,Krebesz2012Implementation}. Furthermore, in the high-rate narrowband wireless communication systems, such as cellular network and WiFi network, there are not many application examples of communication using chaos in the existing literature. The main reasons lie in three aspects: first, the coherent chaos communications between the transmitter and the receiver is difficult to be achieved because the chaos synchronization under a multipath time-varying channel is difficult; second, the non-coherent chaos communications, like DCSK and its variants, need wide frequency spectrum, which is not suitable for the band-limited wireless channel; third, the generation of chaotic signals, using the impulse control \cite{Ren2016Chaos} in the existing chaotic dynamic modulation method, requires complicated hardware and software support.

To address these three issues for chaos-based wireless communications, we propose in this paper the use of chaos as the baseband waveform and a sinusoidal signal as the carrier waveform in the narrowband wireless communication. The generation of a chaotic waveform with low complexity is the priority. A method for constructing a continuous-time chaotic waveform was proposed by using linear superposition of a special basis function \cite{Hayes2005Chaos}; the sufficient conditions for the basis function ensured that the continuous-time waveform was topologically conjugate to the symbolic dynamics (message)\cite{Hirata2005Constructing}. However, the proposed basis function contains an infinitely long and exponentially increasing oscillations, which is non-causal and impractical to be realized. A hybrid chaotic system and its implemented circuit were proposed in \cite{Corron2010}; the output waveform had positive Lyapunov exponents and could be treated as a linear convolution of a symbol sequence and a fixed basis function. The corresponding matched filter could be used to avoid the complicated chaos synchronization between the transmitter and the receiver. For wireless communication using chaos, theoretical analyses have shown that the topological entropy of the hybrid chaotic system is unaffected by the wireless multipath propagation \cite{Ren2013}. The decoding method using a chaotic property was proposed in \cite{Yao2017PRE}, which showed reasonable multipath resistance under the time-varying multipath channel. However, the chaotic signal generated by the hybrid chaotic system cannot be used directly in practiced wireless channels because of limited bandwidth \cite{Zhu2003Numerical} in the conventional wireless communication systems. To address this problem, the idea of using the chaotic signal as communication baseband signal was proposed in \cite{Ren2018Radio}, and a radio-wave wireless communication system with chaos was built based on the channel simulator.

The results in \cite{Ren2018Radio} suggested a simple laboratory demonstration for the wireless communication with chaos, but the theoretical basis of the chaotic encoding and the realization process of the field experiment have not been undertaken so far. In this work, a wireless communication experimental system using a chaotic waveform is being proposed. First, the chaotic waveform, generated from the hybrid dynamical system \cite{Corron2010}, is proved to be topologically conjugate to the symbolic dynamics, thus the symbol sequence (information) is decodable from the chaotic waveform. Second, a chaotic shaping filter is implemented by using a Finite Impulse Response (FIR) filter to encode arbitrary information sequence into the chaotic waveform. At the receiver, the matched filter corresponding to the chaotic waveform is used to maximize the Signal-to-Noise Ratio (SNR), and the method in \cite{Yao2017PRE} is used to relieve the ISI. Finally, experimental tests are carried out using the Wireless open-Access Research Platform (WARP) \cite{WARPweb} for the practical wireless channel. The contributions of this paper are summarized as follows.

\hangafter=1
\setlength{\hangindent}{2.3em}
1)   The chaotic signal is used as a communication baseband waveform and the sinusoidal signal is used as the carrier waveform. Comparing with the traditional communication methods using chaotic signal as spread sequence, the bandwidth efficiency is higher and suitable for the band-limited and high-rate wireless communication applications, which avoids the main obstacle of applying chaos in a narrowband wireless channel. Comparing with the existing works using sinusoidal signal as carrier waveform and the chaotic signal being transmitted in each narrowband subcarrier, the proposed method uses chaotic signal as baseband waveform, and chaotic properties, such as simple matched filter structure \cite{Corron2015Chaos} and multipath ISI resistance, can be used to improve the performance.

\hangafter=1
\setlength{\hangindent}{2.3em}
2) At the transmitter, the chaotic waveform in our method is formulated by the convolution of  the basis function with given symbolic sequence (information), which is implemented by using an FIR filter. This encoding method is simpler as compared with the impulse control method \cite{Ren2016Chaos} that needs a complicated hardware and software support.

\hangafter=1
\setlength{\hangindent}{2.3em}
3) At the receiver, the matched filter corresponding to the chaotic waveform is used to maximize the SNR, which avoids the complicated chaos synchronization between the transmitter and the receiver. The filter output signal and the threshold determined using a chaos property are employed to decode information, which relieves the multipath effects without the traditional complicated channel equalization, and it achieves better BER performance as compared to the traditional method even with the equalization.

\hangafter=1
\setlength{\hangindent}{2.3em}
4) Both encoding and decoding algorithms are compatible with traditional systems. The proposed chaos-based wireless communication scheme is easily implementable in the universal wireless communication platform, improving the performance without modification to the hardware configuration and software structure.

This paper is organized as follows. The hybrid chaotic system and its statistical properties are given in Sec. II. The encoding and decoding algorithms of the proposed method are explained in details in Sec. III. The theoretical performance of the proposed method is analyzed in Sec. IV. Experimental performances of both the proposed method and the conventional non-chaotic method are evaluated in Sec. V. Finally, some conclusions are given in Sec. VI.

\section{Hybrid chaotic system and its statistical properties}
A hybrid dynamical system \cite{Corron2010} given by
\begin{equation}\label{Dynamic}
\ddot{x}(t)-2\beta\dot{x}(t)+(\omega^2+\beta^2)x(t)=(\omega^2+\beta^2)s(t) \\
\end{equation}
generates chaotic signal, $x(t)$ \cite{Corron2010,Ren2016Chaos}, and the parameters are $\omega=2{\pi}$ and $0<\beta{\le}$ln2. $\ddot{x}(t)$ and $\dot{x}(t)$ represent the two-order and first-order differential terms of $x(t)$ with respect to time variable $t$. $s(t)={\rm{sgn}}(x(t))$ switches its value when $\dot{x}(t)=0$, keeping its value at other times. The generated chaotic signal has exact analytic solution given by
\begin{equation}\label{xt}
x(t)=\sum_{m=-\infty}^{\infty}s_m{\cdot}p(t-m),
\end{equation}
where ${s_m}$ is the bi-polar information symbol. $p(t)$ in Eq. (\ref{xt}) is the basis function given by
\begin{equation}\label{pt}
p(t)=\left\{
\begin{aligned}
&(1-e^{-\beta})e^{{\beta}t}({\rm{cos}}{\omega}t-\frac{\beta}{\omega}{\rm{sin}}{\omega}t),\ (t<0) \\
&1-e^{\beta(t-1)}({\rm{cos}}{\omega}t-\frac{\beta}{\omega}{\rm{sin}}{\omega}t),\ \ (0\le{t}< 1)\\
&0,\qquad \qquad \qquad \qquad \qquad \qquad (t\ge 1),
\end{aligned}
\right.
\end{equation}
\subsection{Topological conjugation between the chaotic waveform and the symbol sequence}
For communication systems, the information symbols can be recovered from the received signals only if the encoding baseband waveform and the symbol sequence are topologically conjugate \cite{Hayes2005Chaos, Hirata2005Constructing}. By defining $\phi_m(t)=s_m{\cdot}p(t-m)$, Eq. (\ref{xt}) can be rewritten as
\begin{equation}\label{phim}
x(t)=\sum_{m=-\infty}^{\infty}s_m{\cdot}p(t-m)=\sum_{m=-\infty}^{\infty}\phi_m(t).
\end{equation}

\begin{figure*}[htbp]
\centering{}\includegraphics[scale=0.9]{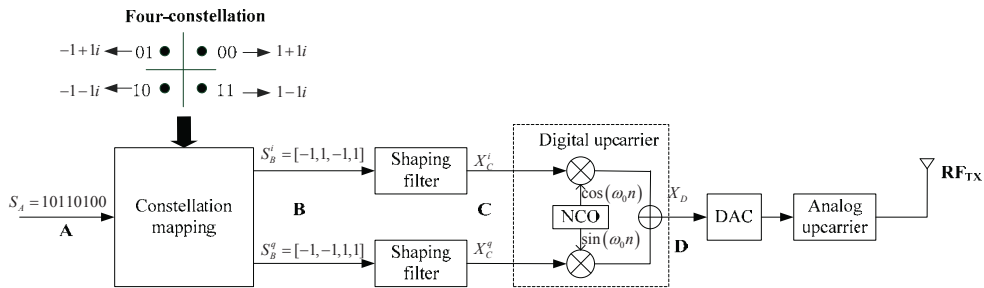}\caption{Schematic diagram of the SDR transmitter}\label{TxBlock}
\end{figure*}

The sufficient conditions for the topological conjugation between chaotic waveform $x(t)$ and driving symbols ${s_m}$ $(m=-\infty,\cdots,\infty)$ are give in \cite{Hirata2005Constructing} as

1) $\phi_k(t)\neq \phi_j(t)$ almost everywhere for $k\neq j$,

2) $\max \limits_{k,j}|\phi_k(t)- \phi_j(t)|<G\cdot2^{-|t|}$, where $G\in\mathbb{R}$,

3) $\min \limits_{k\neq j}\int_{0\leq t<1}|\phi_k(t)- \phi_j(t)|{\rm d}t>\int_{t<0}\max \limits_{k,j}|\phi_k(t)- \phi_j(t)|{\rm d}t$.

Under the assumptions of $s_m\in\{-1,1\}$ and $\beta={\rm ln2}$, the above conditions are satisfied. The proof is as follows:

For condition 1), $\phi_k(t)\neq \phi_j(t)$ for $k\neq j$ is satisfied, because $|p(t-k)|\neq |p(t-j)|$  when $s_k\in\{-1,1\}$, $s_j\in\{-1,1\}$.

For condition 2), we have
\begin{equation}\label{condition2}
\begin{aligned}
&\max \limits_{k,j}|\phi_k(t)- \phi_j(t)|\\
=&\max \limits_{k,j}|s_k\cdot p(t-k)-s_j\cdot p(t-j)|\\
\leq &2\cdot|p(t)|\\
<&5\cdot (e^{\beta})^{-|t|}.
\end{aligned}
\end{equation}
Then the condition 2) is satisfied when $G\ge5$ and $\beta={\rm ln2}$.

For condition 3), the inequality is satisfied for $\beta={\rm ln2}$, because
\begin{equation}\label{condition3}
\left\{
\begin{aligned}
\min \limits_{k\neq j}\int_{0\leq t<1}|\phi_k(t)- \phi_j(t)|{\rm d}t&=2\int_{0\leq t<1}|p(t)|{\rm d}t\approx2.4312\\
\int_{t<0}\max \limits_{k,j}|\phi_k(t)- \phi_j(t)|{\rm d}t&=2\int_{t<0}|p(t)|{\rm d}t\approx0.4642.
\end{aligned}
\right.
\end{equation}
End of proof.

The significance of the topological conjugation lies in that we can not only construct the chaotic waveform from the symbol sequence, but also recover the symbol sequence from the chaotic waveform conversely.

\subsection{Statistical properties of the chaotic waveform over wireless channel}
The chaotic waveform generation using the hybrid chaos system (given by Eq.(\ref{Dynamic}) and Eq. (\ref{phim})) has been proved to be a forward-time dynamics \cite{Corron2010}, who has three Lyapunov exponents (LEs) $\lambda_1=\beta$, $\lambda_2=0$ and $\lambda_3<-\beta$. For wireless communication applications\cite{Ren2013}, the wireless channel is assumed to have multipath with different damping and limited bandwidth modeled by a linear filter. If $\lambda_h$ represents the LE of the channel filter, $\lambda_h<0$, the wireless channel effect on a chaotic signal is the addition of this negative LE and many negative infinite LEs to the original signal, and the positive LE remains unchanged \cite{Ren2013}. In practice, the value of $|\lambda_h|$ is proportional to the channel bandwidth.

If $\bm\lambda_T=\{\lambda_3,\lambda_2,\lambda_1\}$ represents the LE spectra of the transmitted chaotic signal and $|\lambda_h|>|\lambda_3|$, then the LE spectra of the received signal over wireless channel is given by
\begin{equation}\label{LEr}
\bm\lambda_R=[\underbrace{-1/{\Delta t},\cdots,-1/{\Delta t}}_{k_L},\lambda_h,\lambda_3,\lambda_2,\lambda_1],
\end{equation}
where $\Delta t$ is a sufficiently small time interval, and $\tau_L=k_L\Delta t$ is the largest propagation delay of the wireless channel. Equation (\ref{LEr}) shows that the received signal contains all of the three LEs of the transmitted chaotic signal. Furthermore, the Lyapunov dimension of the chaotic signal is unchanged by the wireless channel if $|\lambda_h|>|\lambda_3|$, which means channel bandwidth is enough for chaotic signal transmission.

These results are very important for the use of chaos in wireless communication, because they imply that there is no loss of information during the waveform transmission. Thus, in theory, we can completely recover the transmitted symbols (information) at the receiver using a proper method.

\begin{figure}[htbp]
\centering{}\includegraphics[scale=0.4]{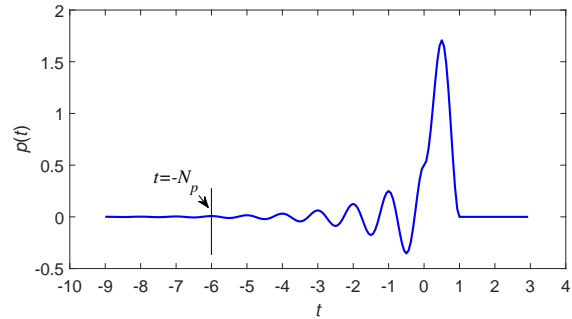}\caption{The basis function $p(t)$ for $\beta=\rm{ln2}$.}\label{BasisFunction}
\end{figure}

\section{Wireless communication system using the chaotic baseband waveform}
In order to be compatible with the traditional non-chaotic wireless communication systems, the Software Defined Radio (SDR) architecture is used. In such a case, the performances of both the chaotic and non-chaotic methods can be evaluated fairly on the same platform. The details of the transmitter and the receiver for the proposed method are explained as follows.

\subsection{Transmitter}
The schematic diagram of SDR transmitter is shown in Fig. \ref{TxBlock}. The signals at the ports A, B, C and D are represented as $S_A$, $S_B$, $X_C$ and $X_D$, respectively. The original information $S_A$ is a binary bit sequence, e.g., $S_A=[10110100]$. In general, there is no grammar constraint on the information, and '0' and '1' have equal probabilities. In modern digital communication, the binary information symbols are usually mapped into the predefined constellation diagram using constellation mapping, such as Multiple Phase Shift Keying (MPSK) and Multilevel Quadrature Amplitude Modulation (MQAM). After the constellation mapping, two signal sequences (in-phase signal $S_B^i$ and quadrature signal $S_B^q$) are obtained at port B. In Fig. \ref{TxBlock}, an example of four-constellation QPSK mappings of $S_A$ is shown, and the mapped signals $S_B^i=[-1,1,-1,1]$ and $S_B^q=[-1,-1,1,1]$.
\begin{figure}[htbp]
\centering{}\includegraphics[scale=0.8]{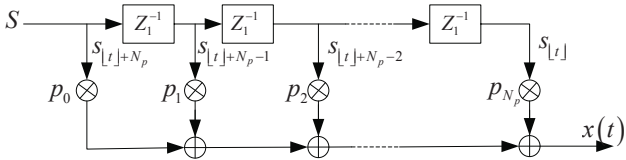}\caption{FIR filter block diagram to implement the shaping filter.}\label{TxFIR}
\end{figure}
The mapped signals, whose waveform are rectangular functions, cannot be transmitted directly because very large bandwidth is needed, which is not available. In practice, a shaping filter is used to convert the rectangular waveform to the baseband signal. The most commonly used filter in traditional communication is the raised cosine filter, which satisfies the Nyquist's ISI-free criterion. The most important distinction between our scheme and the traditional communication lies both in the baseband waveform and the corresponding shaping filter. In the following, the shaping filter used in our method is described.

From the descriptions in Section II, we know that the chaotic waveform can be generated using linear superposition in Eq. (\ref{xt}) of the basis function and the information symbol. In Eq. (\ref{xt}), an infinite symbol sequence is required for all $|t|<\infty$, this is infeasible in practical communication system. Figure \ref{BasisFunction} gives the basis function in Eq. (\ref{pt}) for $\beta=\rm{ln2}$. It can be seen that $p(t)\approx0$ when $t\geq1$, then Eq. (\ref{xt}) can be approximately rewritten as
\begin{equation}\label{xtapprox}
x(t)=\sum_{m=\lfloor t \rfloor}^{\lfloor t \rfloor+N_p}s_m{\cdot}p(t-m),
\end{equation}
where $\lfloor t \rfloor$ indicates the largest integer less than or equal to $t$, and $N_p$ is a positive integer satisfying $p(t)\approx0$ for $t<-N_p$; Eq. (\ref{xtapprox}) reveals that $x(t)$ only depends on the current and future $N_p$ symbols.  From Fig. \ref{BasisFunction}, for $N_p=6$ the requirement is satisfied. By this way, the shaping filter of the proposed method can be implemented by using the FIR filter. The block diagram of the FIR filter used to implement the shaping filter is shown in Fig. \ref{TxFIR}, where symbol sequence $S$ is the input signal. The unit delay, indicated by $z_1^{-1}$, is set as the symbol period of $S$, the taps number is $N_p+1$ and the taps coefficients  $p_0, p_1, \cdots, p_{N_p}$ are determined by the basis function, given as
\begin{equation}\label{pn}
p_n=p(t-\lfloor t \rfloor-N_p+n),\qquad n=0,1,\cdots,N_p.
\end{equation}
Usually, the sampling rate of the shaping filter output should be larger than the symbol rate of input signal. Assuming that the symbol rate at port B is $r_B$ and the sampling rate at port C is $r_C$, we define
\begin{equation}\label{Nc}
N_C=\frac{r_C}{r_B}
\end{equation}
as the oversampling rate, and $N_C>=2$. Then the sampling filtered signal at port C is
\begin{equation}\label{Xc}
X_C=[x_C(0),x_C(1),x_C(2),\cdots],
\end{equation}
where the $n$th element $x_C(n)=x(n/N_C)$.

\begin{figure}[tb]
\centering{}\includegraphics[scale=0.35]{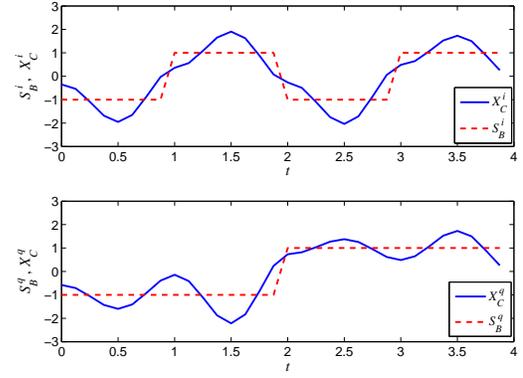}\caption{The chaotic waveform using FIR filter ($N_C=8$). }\label{ChaoticWaveform}
\end{figure}
\begin{figure*}[htbp]
\centering{}\includegraphics[scale=1.1]{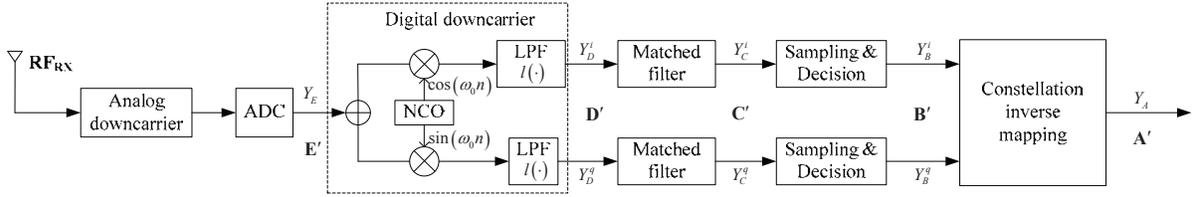}\caption{Schematic diagram of the SDR receiver}\label{RxBlock}
\end{figure*}
When $N_C=8$, the plots of symbol sequences, $S_B^i=[-1,1,-1,1]$ and $S_B^q=[-1,-1,1,1]$, at port B and their corresponding filter output (chaotic waveform), $X_C^i$ and $X_C^q$, at port C are shown in Fig. \ref{ChaoticWaveform}. Because the symbol sequences and the chaotic waveform satisfy the topologically conjugate conditions as given in section II, we can retrieve the symbol sequences by sampling the filter output signal and comparing with a predetermined threshold.

After the shaping filter, a digital upcarrier is used to combine the in-phase and quadrature signals together, as shown in Fig. \ref{TxBlock}. The digital carrier signals, $\cos(\omega_0n_1)$ and $\sin(\omega_0n_1)$, are generated by the Numerically Controlled Oscillator (NCO) given by
\begin{equation}\label{NCO}
\left\{
\begin{aligned}
\cos(\omega_0n_1)=\cos(2\pi\frac{f_b}{f_s}n_1)\\
\sin(\omega_0n_1)=\sin(2\pi\frac{f_b}{f_s}n_1)\\
\end{aligned}
,\qquad (n_1=0,1,2,\cdots),
\right.
\end{equation}
where $f_b$ is the digital carrier frequency, and $f_s$ is the sampling frequency of the digital carrier. The value of $f_s$ is depended on the converting speed of Digital-to-Analog Convert (DAC). The signal at port D is
\begin{equation}\label{Xd}
X_D=[x_D(0),x_D(1),x_D(2),\cdots],
\end{equation}
where the $n_1$ element is given as $x_D(n_1)=x_C^i(n_1)\cos(\omega_0n_1)+x_C^q(n_1)\sin(\omega_0n_1)$, $x_C^i(n_1)$ and $x_C^q(n_1)$ are the in-phase and the quadrature signals at port C, respectively.

In wireless communication systems, a DAC is used to convert the digital signal to analog signal, and a analog upcarrier is used to shift the signal to a higher frequency signal, which is converted into the electromagnetic wave by the antenna ${\rm RF_{TX}}$ and transmitted over the wireless media.

It is worth noting that Fig. \ref{TxBlock} is a  general transmitter structure, which can be used for both the proposed chaotic system and the traditional non-chaotic system. The only difference between two transmitters is the shaping filter, but the other parts are exactly the same.

\subsection{Receiver}
The schematic diagram of SDR receiver is shown in Fig. \ref{RxBlock}, which is a reverse process of the transmitter in Fig. \ref{TxBlock}. The signals at the ports A$'$, B$'$, C$'$, D$'$ and E$'$ are represented as $Y_A$, $Y_B$, $Y_C$, $Y_D$ and $Y_E$. After the wireless propagation and analog downcarrier, the analog RF signal received by antenna ${\rm RF_{RX}}$ is converted to digital signal using Analog-to-Digital Converter (ADC). Thus, $Y_E=[y_E(0), y_E(1), y_E(2), \cdots] $ is a sampling signal with the digital carrier frequency $\omega'_0$. In order to detect the information symbols, the digital carrier should be removed to get the digital baseband signal. In Fig. \ref{RxBlock}, the digital downcarrier and the Low Pass Filter (LPF) are used to remove the digital carrier. Ideally, $\omega'_0$ is equal to the NCO's output frequency $\omega_0$, and so the digital carrier can be removed completely. We assume that there is no frequency offset in the signal $Y_E$ and that $l(\cdot)$ is the transfer function of LPF; the digital baseband (in-phase and quadrature) signals at port D$'$ are
\begin{equation}\label{YD}
\left\{
\begin{aligned}
y_D^i(n)=l(y_E(m_1)\cos(\omega_0m_1))\\
y_D^q(n)=l(y_E(m_1)\sin(\omega_0m_1))\\
\end{aligned}
,\qquad (m_1=0,1,2,\cdots),
\right.
\end{equation}
where $y_E(m_1)$ is the $m_1$th element of $Y_E$.

\begin{figure}[htb]
\centering{}\includegraphics[scale=0.75]{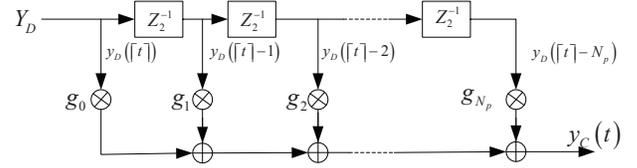}\caption{FIR filter block diagram to implement the matched filter.}\label{RxFIR}
\end{figure}

In noisy environment, a matched filter for a given waveform is the optimal filter for detecting the waveform \cite{Digitalcommunicaitons2008}. For the chaotic waveform generated by Eq. (\ref{xtapprox}), the matched filter can be realized by
\begin{equation}\label{YC}
y_C(t)=\sum_{m=\lceil t \rceil-N_p}^{\lceil t \rceil}y_D(m)g(t-m),
\end{equation}
where $g(t)=p(-t)$ is the time-reversed of basis function of $p(t)$, $\lceil t \rceil$ indicates the least integer larger than or equal to $t$. Equation (\ref{YC}) reveals that $y_C(t)$ depends on the current and past $N_p$ symbols. Equation (\ref{YC}) can also be implemented by FIR filter structure given in Fig. \ref{RxFIR}.

The taps number in Fig. \ref{RxFIR} is $N_p+1$ and the taps coefficients $g_0, g_1, \cdots, g_{N_p}$ are given by
\begin{equation}\label{gm}
g_m=g(t-\lceil t \rceil+m),\qquad m=0, 1, \cdots, N_p.
\end{equation}

For detecting the symbols at the receiver, the symbol rates $r_B$ at port B$'$ should be the same as that at port B in Fig. \ref{TxBlock}. By sampling $y_c(t)$ at rate $r_B$ and comparing it with a preset threshold $\theta_n$, the symbol corresponding to the $n$th sample can be decoded. We will give a detailed description about determining $\theta_n$ in the next Section. Finally, the binary symbol sequence can be retrieved by constellation inverse mapping.

In this section, a wireless communication system using chaotic waveform is designed in detail. Distinguished from the spread spectrum modulation in the conventional  chaos-based wireless communication systems, the proposed method adopts a novel chaotic shaping filter to generate the chaotic baseband signal. Unlike the existing chaotic symbolic dynamics modulation \cite{Hayes1993Communicating} \cite{Ren2016Chaos}, which requires high precision hardware and complex control algorithm to generate the chaotic waveform, the proposed method uses digital FIR filters to implement the chaotic shaping filter at the transmitter. At the receiver, the matched filter corresponding to the transmitted chaotic waveform is used to maximize the SNR and facilitate decoding the information symbols.

Compared with the coherent chaotic communication system, which needs chaos synchronization to decode information, the proposed method belongs to a non-coherent communication system, since it does not need chaos synchronization. However, this does not mean that the general synchronization is not needed in the proposed method. In our experiment, the general synchronization is used to correct the clock drift at the receiver and to determine the start of a data frame, by using the training bits. In fact, the general synchronization is essential for most of wireless communication systems; the method to achieve this general synchronization in the proposed system is the commonly used one.

These features of the proposed method are very important for the practical communication application of chaos because: 1) it can be implemented on the existing hardware, without changing the transmitter/receiver software structure; 2) the performance comparison with the traditional non-chaotic methods can be done straightforwardly by using the same hardware platform in a just way.

Discussion 1. The proposed chaotic waveform can also be used in the conventional CSK and DCSK systems; for instance, different chaotic generators in the CSK system can be implemented using the different initial values in Eq.(\ref{Dynamic}), and the waveform can be used as the reference signal in the DCSK system \cite{Ren2017A,Bai2018Chaos}.

\section{Performance analysis}
\begin{figure}[htb]
\centering{}\includegraphics[scale=0.68]{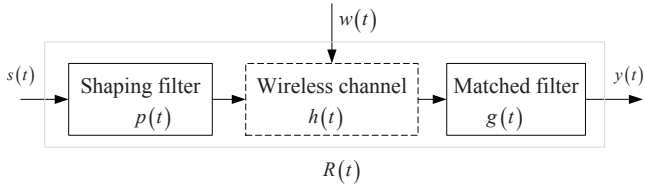}\caption{Simplified block diagram of the wireless communication system using chaotic waveform.}\label{SimplifiedBlockDiagram}
\end{figure}

The BER performance and the computational complexity of the proposed scheme are analyzed in this section. A simplified block diagram (considering the baseband signal transmission only) of the proposed system is shown in Fig. \ref{SimplifiedBlockDiagram}, where $w(t)$ is an Additive White Gaussian Noise (AWGN). The main differences between our proposed system and the traditional non-chaotic system are the shaping filter, the corresponding matched filter and the symbol decision algorithm. $p(t)$ and $g(t)$ are the impulse responses of the shaping filter and the matched filter, which are given by Eq. (\ref{xt}) and Eq. (\ref{YC}), respectively.

The impulse response of the wireless multipath channel is given by
\begin{equation}\label{ht}
h(t)=\sum_{l=0}^{L-1}\alpha_l\delta(t-\tau_l),
\end{equation}
where $\alpha_l$ and $\tau_l$ are the attenuation and propagation delay corresponding to path $l$ from the transmitter to the receiver, and $\delta(\cdot)$ is the Dirac delta function. Assume that delay $\tau_l\ (l=0,1,\dots,L-1)$ satisfies $0=\tau_0<\tau_1<\cdots<\tau_{L-1}$, then the channel fading $\alpha_l$ can be modeled as a negative exponential decay $\alpha_l=e^{-\gamma\tau_l}$, where $\gamma$ is the damping coefficient. Equation (\ref{ht}) is a statistical average channel model \cite{WINNERII2007} for a practical wireless communication channel, and it is considered for the theoretical performance analysis.

\subsection{BER performance}
In Fig. \ref{SimplifiedBlockDiagram}, the total impulse response from the input to output can be described as
\begin{equation}\label{Rt}
\begin{aligned}
R(t)&=p(t)*h(t)*g(t)\\
&=p(t)*\big(\sum_{l=0}^{L-1}\alpha_l\delta(t-\tau_l)\big)*g(t)\\
&=\sum_{l=0}^{L-1}\alpha_l(p(t)*g(t))*\delta(t-\tau_l)\\
&=\sum_{l=0}^{L-1}r(t-\tau_l)
\end{aligned}
,
\end{equation}
where '*' denotes convolution and $r(t-\tau_l)$ can be calculated from Eq. (\ref{rtl}), in which $A=\frac{(\omega^2-3\beta^2)}{4\beta(\omega^2+\beta^2)}$, $B=\frac{(3\omega^2-\beta^2)}{4\omega(\omega^2+\beta^2)}$ and $D=e^{-\beta|t-\tau_l|}$.
\begin{figure}[htbp]
\centering{}\includegraphics[scale=1]{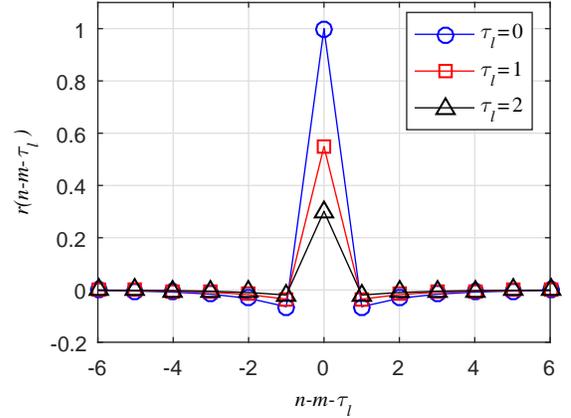}\caption{($\gamma=0.6$) plots of $r(n-m-\tau_l)$ for different $\tau_l$.}\label{AutoCorr}
\end{figure}
\begin{equation}\label{rtl}
\begin{aligned}
&r(t-\tau_l)=\\
&\left\{
\begin{aligned}
\alpha_lD(2-e^{-\beta}-e^{\beta})\big(A\cos\omega(t-\tau_l)+B\sin\omega(t-\tau_l)\big),\ \ \ \ \\
    {\rm{if}}\ \ (|t-\tau_l|\ge1)\ \ \ \ \\
\alpha_l
\left\{
    \begin{aligned}
    &A\big(D(2-e^{-\beta})-D^{-1}e^{-\beta}\big)\cos\omega(t-\tau_l)+ \\
    &B\big(D(2-e^{-\beta})+D^{-1}e^{-\beta}\big)\sin\omega(t-\tau_l)+1-|t-\tau_l|
    \end{aligned}
    \right\}.\\
    {\rm{if}}\ \ (0\le|t-\tau_l|<1)\ \ \ \
\end{aligned}
\right.
\end{aligned}
\end{equation}

The output signal $y(t)$ is
\begin{equation}\label{yt}
\begin{aligned}
y(t)&=s(t)*R(t)+w(t)*g(t)\\
&=\sum_{m=-\infty}^{\infty}s(m)R(t-m)+W(t)\\
&=\sum_{l=0}^{L-1}\sum_{m=-\infty}^{\infty}s(m)r(t-m-\tau_l)+W(t)
\end{aligned}
,
\end{equation}
where $s(m)\in\{-1,1\}, m=-\infty, \cdots, \infty$ is a bi-polar symbol sequence, as given at port B in Fig. \ref{TxBlock}, with symbol rate $r_B$. $W(t)=w(t)*g(t)$ is the filtered noise. If $w(t)$ is an AWGN with zero mean, then $W(t)$ is also Gaussian noise with zero mean \cite{Blakely2013Communication}. $y(t)$ is the output signal of the matched filter, corresponding to the signal at port C$'$ in Fig. \ref{RxBlock}.

By sampling $y(t)$, using the interval $t=1/r_B$ and recoding the $n$th sampling value as $y(n)$, we have
\begin{equation}\label{yn}
\begin{aligned}
&y(n)=\sum_{l=0}^{L-1}\sum_{m=-\infty}^{\infty}s(m)r(n-m-\tau_l)+W(n)\\
&=\underbrace{s(n)\sum_{l=0}^{L-1}r(\tau_l)}_{Expected\ signal}+\underbrace{\sum_{\substack{m \ne {n} \\ m=-\infty}}^{m=\infty}s(m)\sum_{l=0}^{L-1}r(n-m-\tau_l)}_{Inter-symbol\  interference}+\underbrace{W(n)}_{Noise}
\end{aligned}
.
\end{equation}

In Eq. (\ref{yn}), the first term contains the expected symbol $s(n)$, the second term is the ISI from other symbol $s(m), m\neq n$. The values of $r(n-m-\tau_l)$ for different $\tau_l$ are plotted in Fig. \ref{AutoCorr}. We can see that the ISI of chaotic communication system comes from two sources: First, considering any one of the lines in Fig. \ref{AutoCorr}, it shows that there exist only several $n-m-\tau_l\neq0$ such that $r(n-m-\tau_l)\neq0$, which means that the communication signal using chaotic baseband waveform does not satisfy the Nyquist's ISI-free criterion \cite{Digitalcommunicaitons2008}, and there is an ISI even in the single path channel, although the ISI in the single path is small. Second, considering all of the three lines in Fig. \ref{AutoCorr}, we have $r(n-m-\tau_l)\neq0$ when $n-m-\tau_l=0$, which means the ISI exists due to the multipath propagation.

To detect the expected symbol, $s(n)$, from the sampled filtered signal, $y(n)$, a threshold $\theta_n$ is used to decode the symbol as given by
\begin{equation}\label{sn}
s(n)=\left\{
\begin{aligned}
1,\ \ \ \ &{\rm if}\ \ (y(n)\geq \theta_n)\\
-1,\ \ \ \ &{\rm if}\ \ (y(n)<\theta_n)
\end{aligned}
.
\right.
\end{equation}

The intuitive threshold is $\theta_n=0$, but a proper threshold can be designed to relief the ISIs from both aforementioned sources \cite{Yao2017PRE}. Under the multipath channel, the optimal threshold is
\begin{equation}\label{OptimalThreshold}
\theta_n^{opt}=\sum_{l=0}^{L-1}\sum_{\substack{m \ne {n} \\ m=-\infty}}^{\infty}s(m)r(n-m-\tau_l),
\end{equation}
in which $r(n-m-\tau_l)$ can be calculated by using Eq. (\ref{rtl}) if the channel parameters $\tau_l$ and $\alpha_l$ are known. Assuming that the variance of $W(n)$ is $\sigma_W^2$ and that $P=\sum_{l=0}^{L-1}r(\tau_l)$ is the total power of the expected symbol $s(n)$, BER using the optimal threshold, $\theta_n=\theta_n^{opt}$, is given as
\begin{equation}\label{BERopt}
    p({\rm{error}}|\theta_n=\theta_n^{opt})=\frac{1}{2}erfc\Bigg(\frac{P}{\sqrt{2\sigma_{W}^{2}}}\Bigg),
\end{equation}
in which $erfc(\cdot)$ is the complementary error function. The optimal threshold $\theta_n^{opt}$ given in Eq. (\ref{OptimalThreshold}) contains both the past symbols $s(m)\ (m<n)$ and the future symbols $s(m)\ (m>n)$. At the current time, the past symbols have been decoded and it can be used for calculating $\theta_n^{opt}$. However, in practice, the future symbols are unknown and $\theta_n^{opt}$ cannot be obtained using the available information. In this case, a suboptimal threshold, using only the past symbols, is given as
\begin{equation}\label{SubOptimalThreshold}
\theta_n^{subopt}=\sum_{\substack{m=-\infty}}^{n-1}s(m)\sum_{l=0}^{L-1}r(n-m-\tau_l).
\end{equation}

Assume that the probabilities of information bits "1" and "-1" are equal, i.e., $pr\big(s(m)=1\big)=pr\big(s(m)=-1\big)=1/2$, where $pr(\cdot)$ is the probability of event '$\cdot$', then the BER using the suboptimal threshold $\theta_n=\theta_n^{subopt}$ is given as
\begin{equation}\label{BERsubopt}
\begin{aligned}
    &p({\rm{error}}|\theta_n=\theta_n^{subopt})\\
     &=\sqrt{2\sigma_{W}^{2}}\frac{e^{\beta}-1}{4|K|}\left\{
    \begin{aligned}
    z_1\cdot erfc(z_1)-z_2\cdot erfc(z_2)\\
    -{e^{-{z_1}^2}}/{\sqrt{\pi}}+{e^{-{z_2}^2}}/{\sqrt{\pi}}\\
    \end{aligned}
    \right\},\\
    \end{aligned}
\end{equation}
where $K=\sum_{l=0}^{L-1}\alpha_l(2-e^{-\beta}-e^{\beta})e^{-\beta\tau_l}\big(Acos(\omega\tau_l)+Bsin(\omega\tau_l)\big)$, $z_1=\big(P+\frac{|K|}{e^{\beta}-1}\big)/{\sqrt{2\sigma_{W}^{2}}}$ and $z_2=\big(P-\frac{|K|}{e^{\beta}-1}\big)/{\sqrt{2\sigma_{W}^{2}}}$.

\begin{figure}[tbh]
\centering{}\includegraphics[scale=0.4]{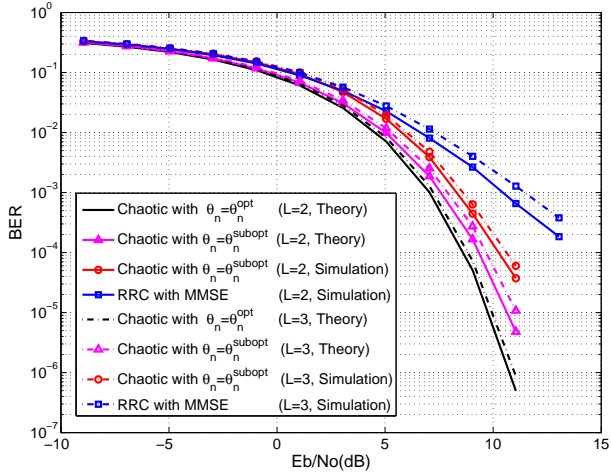}\caption{BER comparisons under static wireless channel.}\label{BER_idealchannel}
\end{figure}

In the two-path channel ($L$=2) case with $\tau_0=0$, $\tau_1=1$ and $\gamma=0.6$, and the three-path channel ($L$=3) case with $\tau_0=0$, $\tau_1=1$, $\tau_2=2$ and $\gamma=0.6$, the BER comparison results between wireless communication system using the proposed chaotic waveform and the conventional non-chaotic waveform are given in Fig. \ref{BER_idealchannel}. For the wireless communication system using chaotic baseband waveform, the theoretical BER results from Eq. (\ref{BERopt}) and Eq. (\ref{BERsubopt}), and the simulation results using the suboptimal threshold $\theta_n=\theta_n^{subopt}$ are given in Fig. \ref{BER_idealchannel} as shown by the legend. For the wireless communication system using non-chaotic baseband waveform, the commonly used Root Raised Cosine (RRC) shaping filter and the corresponding matched filter are used, at the same time, the Minimum Mean Square Error (MMSE) algorithm \cite{Tuchler2002Minimum} is used for the channel equalization. During the simulation, the channel parameters, $\tau_l$ and $\gamma$, are invariant (static channel), and are known by the receiver. All the simulation results are obtained by averaging over 500 000 trials. Figure \ref{BER_idealchannel} shows that the wireless communication system using chaotic baseband waveform and the optimal threshold has the lowest BER. The wireless communication system with the conventional RRC and MMSE has the worst BER. This phenomena is a consequence of a fundamental property of chaotic dynamics , namely the invariant Lyapunov spectrum of the chaotic signal over the multipath (wireless) channel \cite{Ren2013}, that can be used to effectively resist multipath interference without enhancing the noise effect \cite{Yao2017PRE}, while the noise is enhanced by the traditional linear equalization\cite{Goldsmith2005Wireless}. By using the suboptimal threshold, the BERs using chaotic baseband waveform are not only lower than those of the conventional non-chaotic baseband waveform, but also close (0.5dB gap for BER=$10^{-3}$) to the results using the optimal threshold. The simulation results using the suboptimal threshold is slightly worse than that of the corresponding theoretical results. This is because the retrieved past symbols are used in the calculation of the suboptimal threshold, and then the decoding error in the past affects the accuracy of the suboptimal threshold.

\begin{figure}[tbh]
\centering{}\includegraphics[scale=0.4]{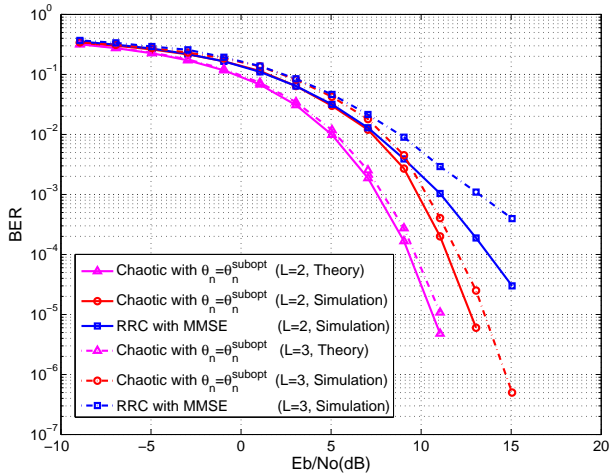}\caption{BER comparisons under time-varying wireless channel.}\label{BER_estchannel}
\end{figure}

In the above analysis, the channel is static during the information transmission and the accurate channel parameters are known for symbol decoding. However, in practice, the wireless channel is time-varying and the channel parameters are unknown in advance. In such case, we consider the wireless channel as a time-varying but quasi-static channel, in which the channel parameter, $\gamma$, is assumed to be unchanged within one data frame, but be variant from one frame to the next. In our simulation, $\gamma$ obeys the uniform distribution in the range of [0.3, 0.9]. Figure \ref{BER_estchannel} gives the BER comparison results under a quasi-static channel. In our simulation, there are 4096 bits in one frame, which contains 256 training bits and 3840 information bits. The training bits are used for channel estimation using the least squares (LS) algorithm. The estimated channel parameters are used to calculate suboptimal threshold of our method and are used as parameters in the MMSE algorithm for the conventional non-chaotic system. The simulation results are obtained by averaging over 500 frames. In Fig. \ref{BER_estchannel}, BER simulation results of the wireless communication system using the proposed chaotic baseband waveform with the suboptimal threshold, the conventional non-chaotic waveform with RRC and MMSE, and the theoretical BER from Eq. (\ref{BERsubopt}) are given for comparison. We know that the simulation BERs of both waveforms are worse than the corresponding results in Fig. \ref{BER_idealchannel}, because of the imperfect channel estimation. However, for both two and three-path channels, the BER performance of the system using the chaotic baseband waveform is better than that of the conventional system using the non-chaotic baseband waveform.

\subsection{Computational complexity}
The computational complexity of the proposed method is analyzed and compared with the traditional non-chaotic method, and some traditional chaotic methods in this subsection. The time complexity, that describes the number of elementary operations needed to perform a method, is used to measure the computational complexity. In general, we focus on the asymptotic behavior of the complexity when the input size increases. The time complexity $T(N)$ is commonly expressed as $T(N)=O(N)$, where $N$ is the input size. To be impartial, the time complexity of all methods for decoding one information bit is analyzed in the following.

 {\linespread{1.5}
 \begin{table*}\label{ComputationComplexityComparisons}
 \centering
   \caption{Computational complexity comparison of four methods}
   \begin{tabular}{c c c c}
    \hline
     \hline
        Method & Shaping filter ($T$) & Matched filter ($T$)  & decoding ($T$)\\
     \hline
    The proposed chaotic method & $O(N_CN_p)$ & $O(N_CN_p)$ & $O(L)$\\
    Traditional non-chaotic method & $O(N_CN_p)$ & $O(N_CN_p)$ & $O(L^3)$\\
    NC-DCSK&$O(N_\beta N_p)$&$O(N_\beta N_p)$&$O(N_\beta)$\\
    NC-CSK&$O(N_\beta N_p/log_2M)$&$O(N_\beta N_p/log_2M)$&$O(MN_\beta/log_2M)$\\
      \hline
      \hline
   \end{tabular}
 \end{table*}}

All the methods contain three main parts: the shaping filter, the matched filter and the decoding algorithm.  The computational complexity comparison of four methods are shown in Table II. For the proposed method, the chaotic shaping filter and the corresponding matched filter are shown in Fig. (\ref{TxFIR}) and Fig. (\ref{RxFIR}), and the input sizes are the taps number and  samples number. If $N_p+1$ is the taps number of filter and $N_C$ is the oversampling rate given in Eq. (\ref{Nc}), then the complexities of both filters are $O(N_CN_p)$. In order to decode a symbol, a proper threshold, $\theta_n$ in Eq. (\ref{OptimalThreshold}) or Eq. (\ref{SubOptimalThreshold}), is computed for symbol decision. The input size in this algorithm is the multipaths number $L$, and the complexity is $O(L)$. For the traditional non-chaotic method, the complexities of the RRC shaping filter and the corresponding matched filter are the same as that in the proposed method, and the complexity of the decoding algorithm, results mainly from the MMSE equalization; it is $O(L^3)$.

For traditional chaotic methods, the MC-DCSK \cite{Kaddoum2013Design} and MC-CSK \cite{Yang2017Multi} are considered. In both methods, multiple symbols are transmitted simultaneously over multiple subcarriers to improve the data rate and the spectral efficiency. The RRC filters are used to reshape the signal waveform corresponding to each subcarrier. In the MC-DCSK method, $M$ subcarriers are used to transmit $M-1$ information bits. The complexities of the shaping filter and the matched filter corresponding to one bit are $O(N_\beta N_p)$, where $N_\beta$ is the spreading factor and $N_p+1$ is the taps number of the filter. The decoding is performed by vector product with complexity $O(N_\beta)$. In the MC-CSK method, $2M$ subcarriers are used to transmit $Mlog_2M$ information bits. Then the complexities of the shaping filter and the matched filter are $O(N_\beta N_p/log_2M)$. The number of vector products in the MC-CSK is $M$ times of that in MC-DCSK, and the complexity is $O(MN_\beta/log_2M)$. The channel equalizations in both methods are not considered, because it is assumed that the path delay is much smaller than the chaotic sequence length $N_\beta$ in one subcarrier.

In practice, the variables mentioned above usually satisfy: $4 \leq N_C \leq 16$, $2 \leq L \leq 6$, $64 \leq N_\beta \leq 256$ and $2 \leq M \leq 64$. We can see that the values of $N_C$ and $L$ are much smaller than the value of $N_\beta$, and the value of $N_C$ is smaller than the value of $N_\beta/log_2M$. Then from Table II, we surmise that the proposed method has less computational complexity than the traditional NC-DCSK, NC-CSK and non-chaotic methods.

\section{Experimental results}
In order to demonstrate the practical performance of the proposed chaotic baseband waveform communication system, a radio-frequency wireless communication platform is used to realize both the chaotic and non-chaotic baseband waveforms communication.

\begin{figure}[hbt]
\centering{}\includegraphics[scale=0.45]{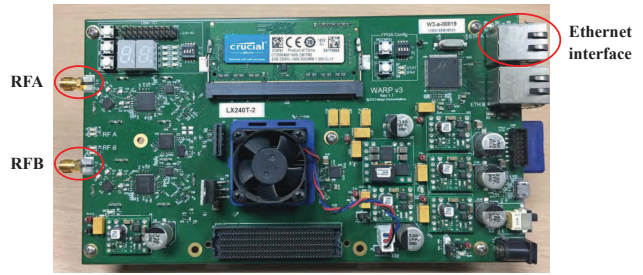}\caption{Photo of the WARP node}\label{WARPNode}
\end{figure}

The experiment is performed using wireless open-access research platform version 3 (WARP V3) designed by Rice University \cite{WARPweb}. The hardware photo of the system is shown in Fig. \ref{WARPNode}, in the WAPR V3, the Xilinx Virtex-6 LX240T FPGA is used for digital signal processing, two MAX2829 RF chips are used to support dual-channel and 2.4GHz/5GHz dual-band transceiver, the maximum transmission power is 20dBm by using the dual-band power amplifier, the 12-bit low power analog/digital converter AD9963 is used to provide two ADC channels with sample rates of 100 MSPS and two DAC channels with sample rates of 170 MSPS, and two 10/100/1000 Ethernet interfaces (Marvell 88E1121R) is used to realize the high-speed digital signal exchange with the Personal Computer (PC).

\begin{figure}[hbt]
\centering{}\includegraphics[scale=1]{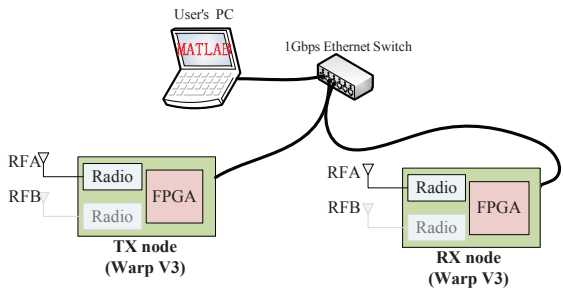}\caption{Block diagram of WARPLab}\label{WARPLab}
\end{figure}

In order to validate the proposed chaotic baseband waveform for point-to-point communication based on the WARP, the WARPLab framework is used. The block diagram of WARPLab is shown in Fig. \ref{WARPLab}, in which two WARP nodes are connected with PC through 1Gbps Ethernet switch. Each WARP node has two radio antennas, which are referred as RFA and RFB. In our test, only the RFAs, operating at the 2.4GHz carrier frequency with 20MHz bandwidth, in both nodes are used as transmitter (TX) and receiver (RX) for Single Input Single Output (SISO) communication.

In our experimental platform, the constellation mapping and the shaping filter are implemented on the PC using MATLAB R2016a, while the digital upcarrier,  DAC and the analog upcarrier are implemented on the WARP transmitter. After wireless propagation, the received RF signal is processed (through analog downcarrier, ADC and digital downcarrier) on the WARP receiver to obtain the complex received signal containing both the in-phase signal and the quadrature signal. The complex signal is then transferred to the PC for matching filtering and symbol detection in MATLAB environment. In this test, the data is transmitted frame by frame. There are 4096 bits in one frame, which contains 1023 training bits (using the Golden sequence) and 3073 data bits. The training bits in our experiment are not only used for channel estimation, but also for frame synchronization and pilot-based frequency offset estimation.

\subsection{The single path channel test}
\begin{figure*}[tbh]
\centering{}
\subfloat[]{\includegraphics[height = 1.3in]{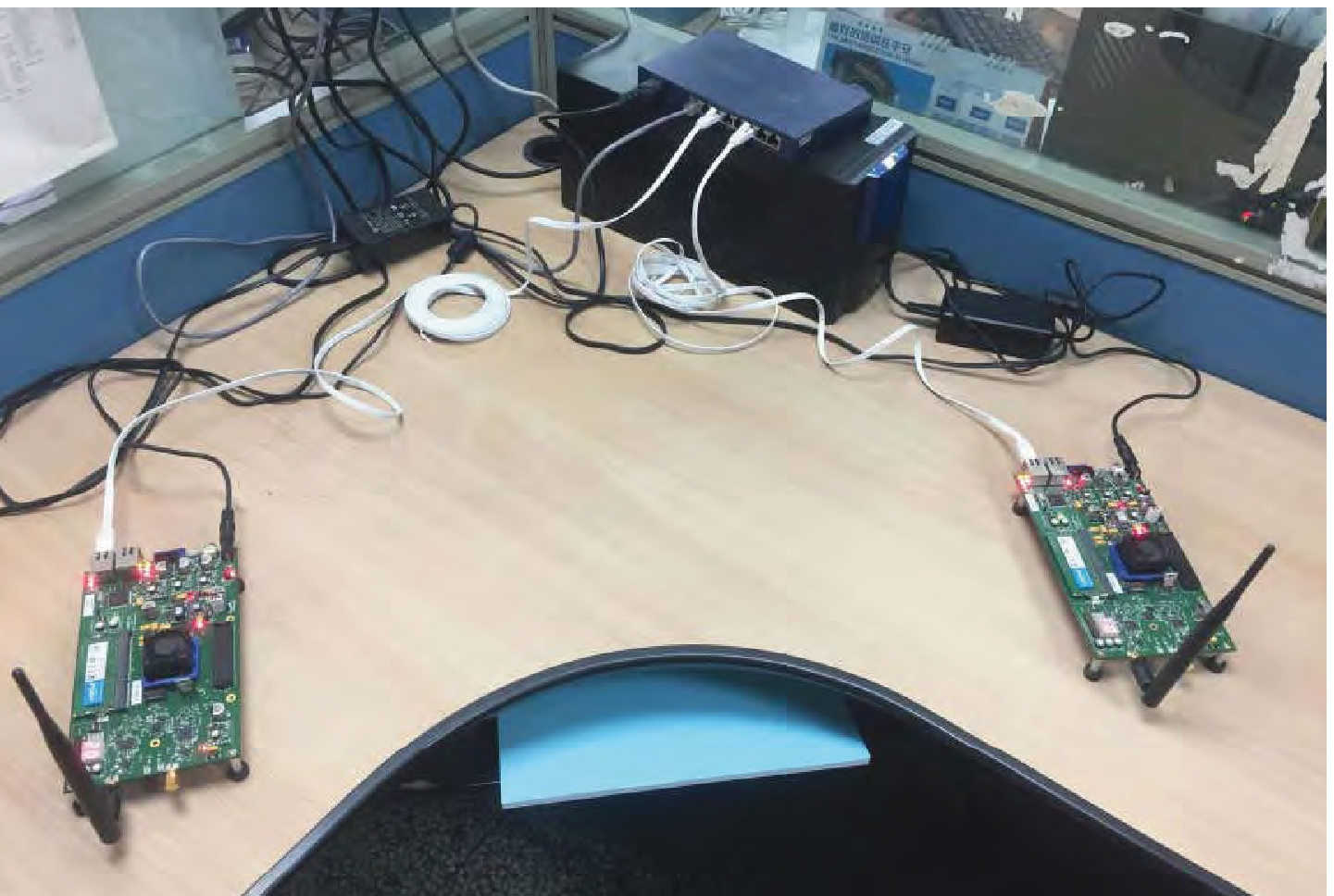}}\ \ \
\subfloat[]{\includegraphics[width=2in, height = 1.3in]{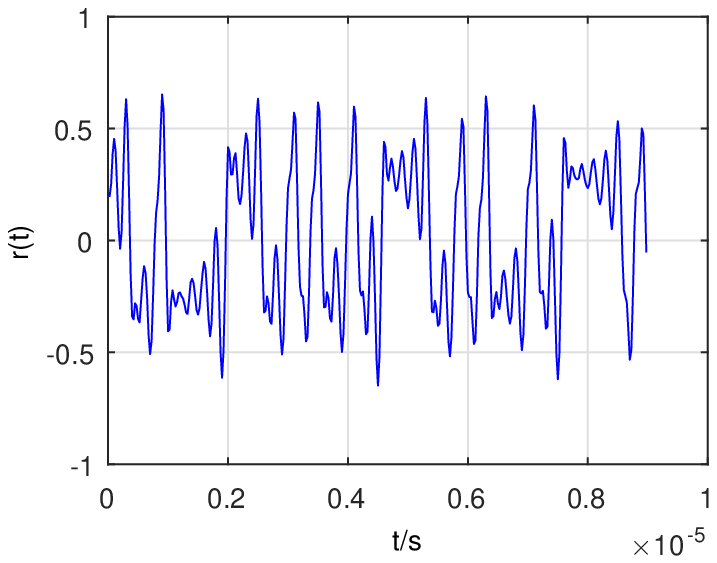}}
\hfil
\subfloat[]{\includegraphics[width=2in, height = 1.3in]{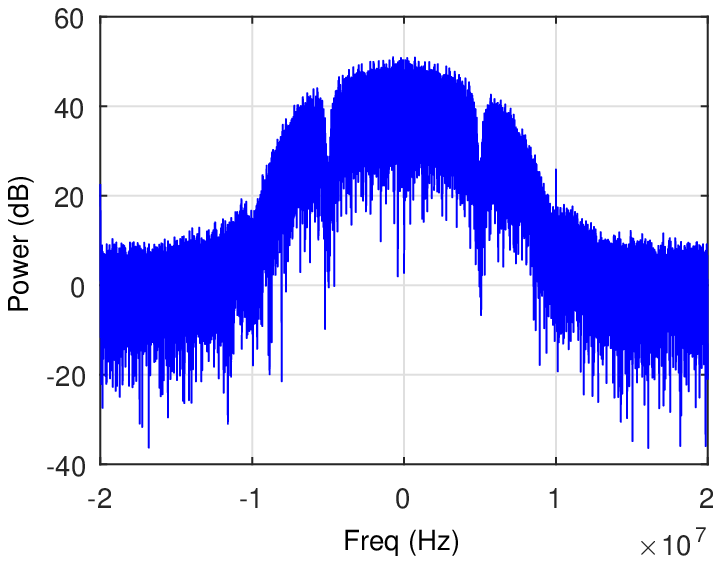}}
\subfloat[]{\includegraphics[width=2in, height = 1.3in]{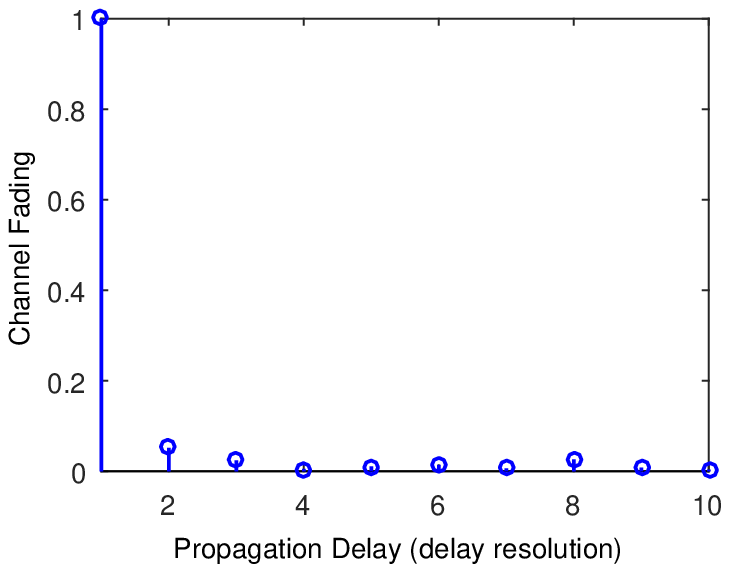}}
\caption{(a) The photo of laboratory test setup, (b) the time domain waveform of received baseband signal, (c) the power spectrum of the received baseband signal, and (d) the normalized estimated channel parameters.}
\label{SinglePathTest}
\end{figure*}

\begin{figure*}[hbt]
\centering{}
\subfloat[]{\includegraphics[height = 1.5in]{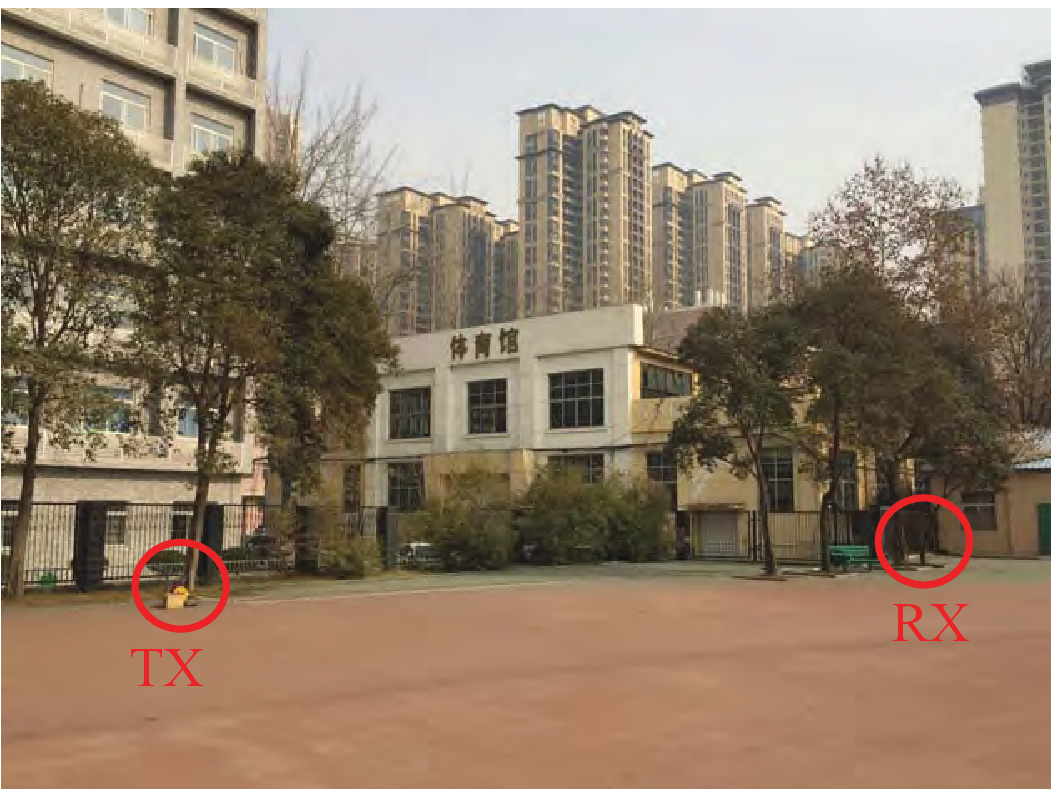}}\
\hfil
\subfloat[]{\includegraphics[height = 1.5in]{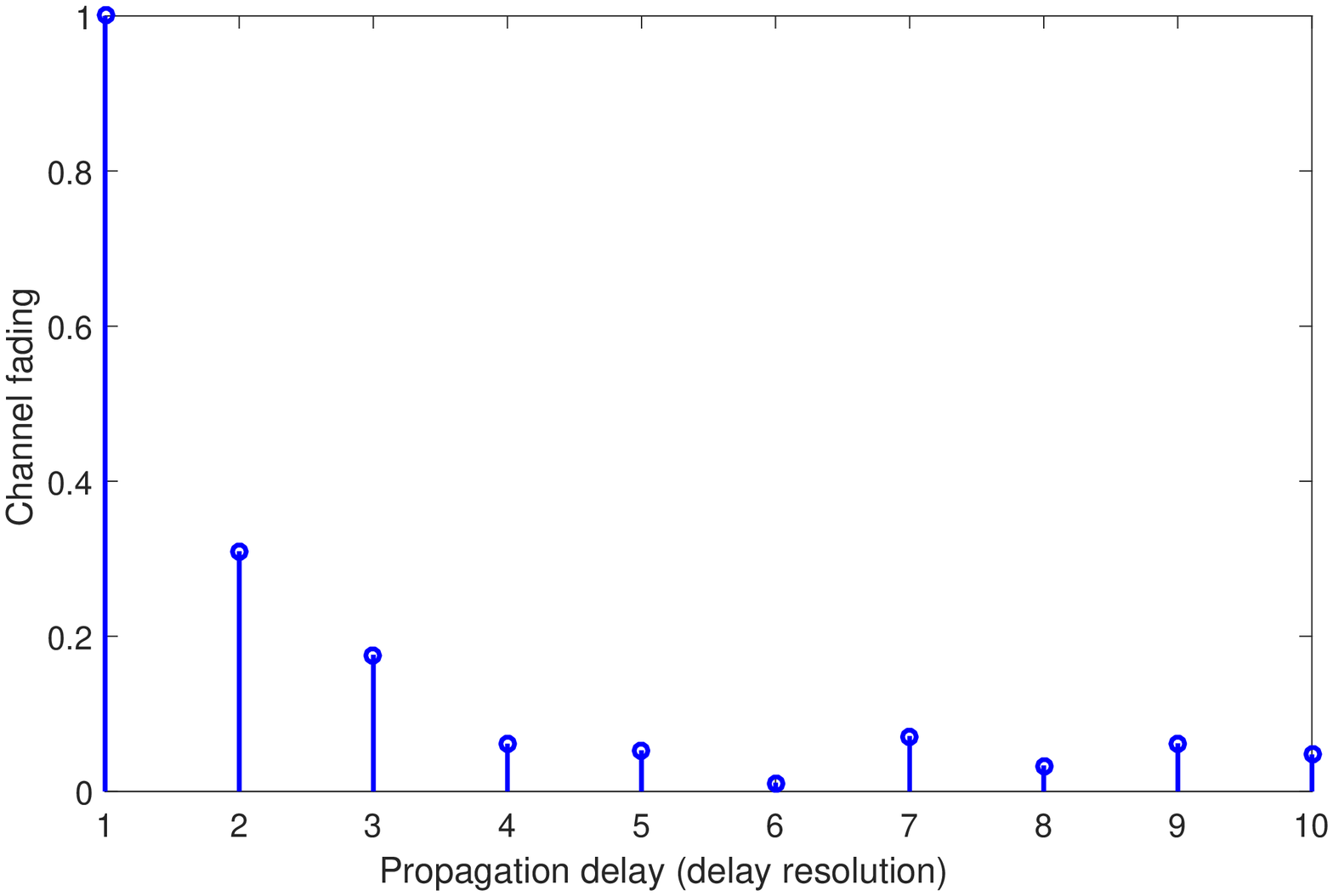}}
\caption{(a) The photo of the test in the first scenario,\ \ \    (b) the normalized estimated channel.}
\label{Scenario1PathTest}
\end{figure*}

At the outset, several basic properties of the experimental platform are tested in the laboratory, and the photo of the test scenario is given in Fig. \ref{SinglePathTest}(a). The chaotic baseband waveform is used and the transmission power is 3.5dBm in this test. After frequency offset calibration, the time domain waveform and the power spectrum of the received baseband signal are shown in Fig. \ref{SinglePathTest}(b) and Fig. \ref{SinglePathTest}(c), respectively. We can see from Fig. \ref{SinglePathTest}(c) that the 10dB bandwidth is about 8 MHz, thus the delay resolution of our experimental system is 0.125$\mu$s. The channel parameters, delay and fading, are estimated by using the LS algorithm, and the normalized estimated channel parameters are shown in Fig. \ref{SinglePathTest}(d). We see that there is one main path with channel fading 1, and the other paths are relatively weak (i.e., the channel fading is less than 0.07). The unit of the x-axis in Fig. \ref{SinglePathTest}(d) is the delay resolution. The received baseband signal is filtered using the corresponding chaotic matched filter, and then the information symbol is decoded. In the laboratory environment, the symbols can be recovered without error. These results prove that the experimental platform can be used for RF wireless communication using chaotic baseband waveform.
\subsection{The multipath channel test}
The BER performance of the proposed method under the real multipath channel is tested, at the same time, the traditional method with the non-chaotic baseband waveform is also tested for comparison. In order to guarantee fairness of comparison, the transmission of both methods should be done as simultaneously as possible, because the channel may be time-varying. In the real environment, the channel coherence time is related to the speeds of terminals and clusters; we verify that the channel does not vary during the transmission of one frame in our test. Then, the 3073 data bits in one frame are divided into two parts, the first 1536 bits are using chaotic baseband waveform and the last 1537 bits are using non-chaotic baseband waveform. The channel parameters, estimated using the LS algorithm and 1023 training bits, are used for both methods. Two tests are carried out at the campus of our university, the first scenario is at the corner of the playground and the second scenario is in the middle of buildings. The experimental results are obtained by averaging over 100 frames.
\begin{figure}[htb]
\centering{}\includegraphics[scale=0.35]{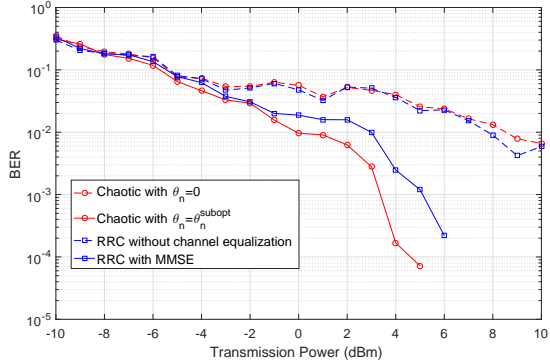}\caption{The experimental BERs comparison in the first scenario}\label{BERFirstScenario}
\end{figure}

\begin{figure*}[htb]
\centering{}
\subfloat[]{\includegraphics[height = 1.8in]{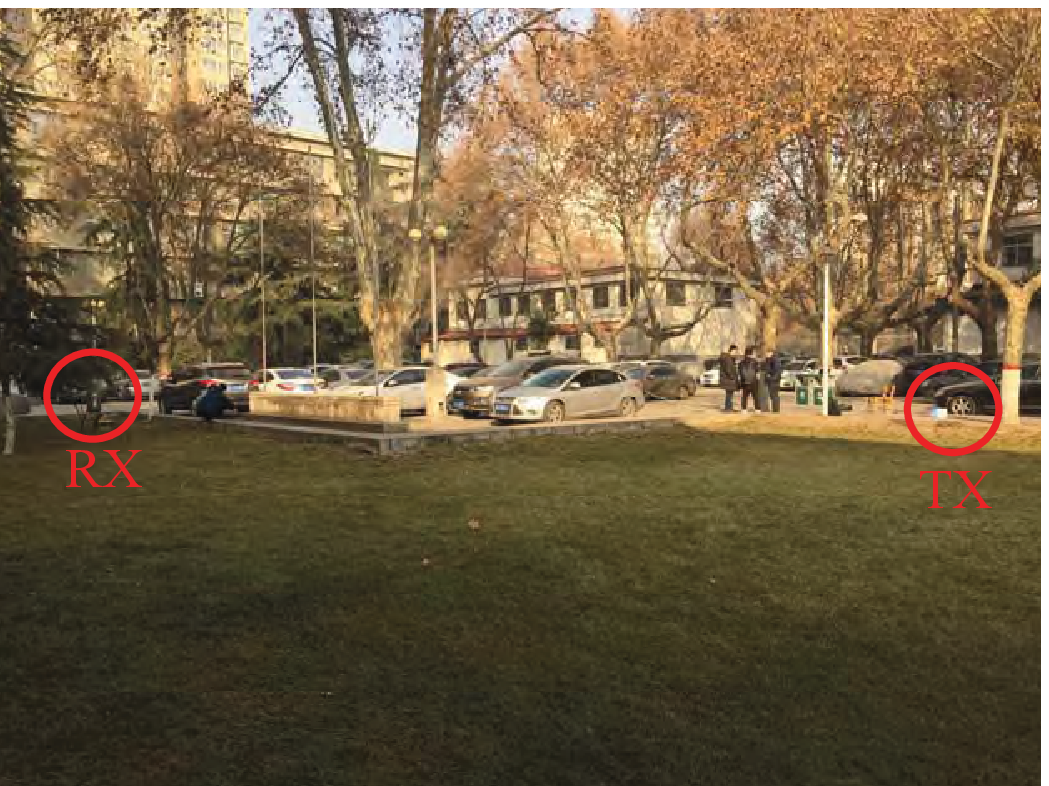}}\
\hfil
\subfloat[]{\includegraphics[height = 1.8in]{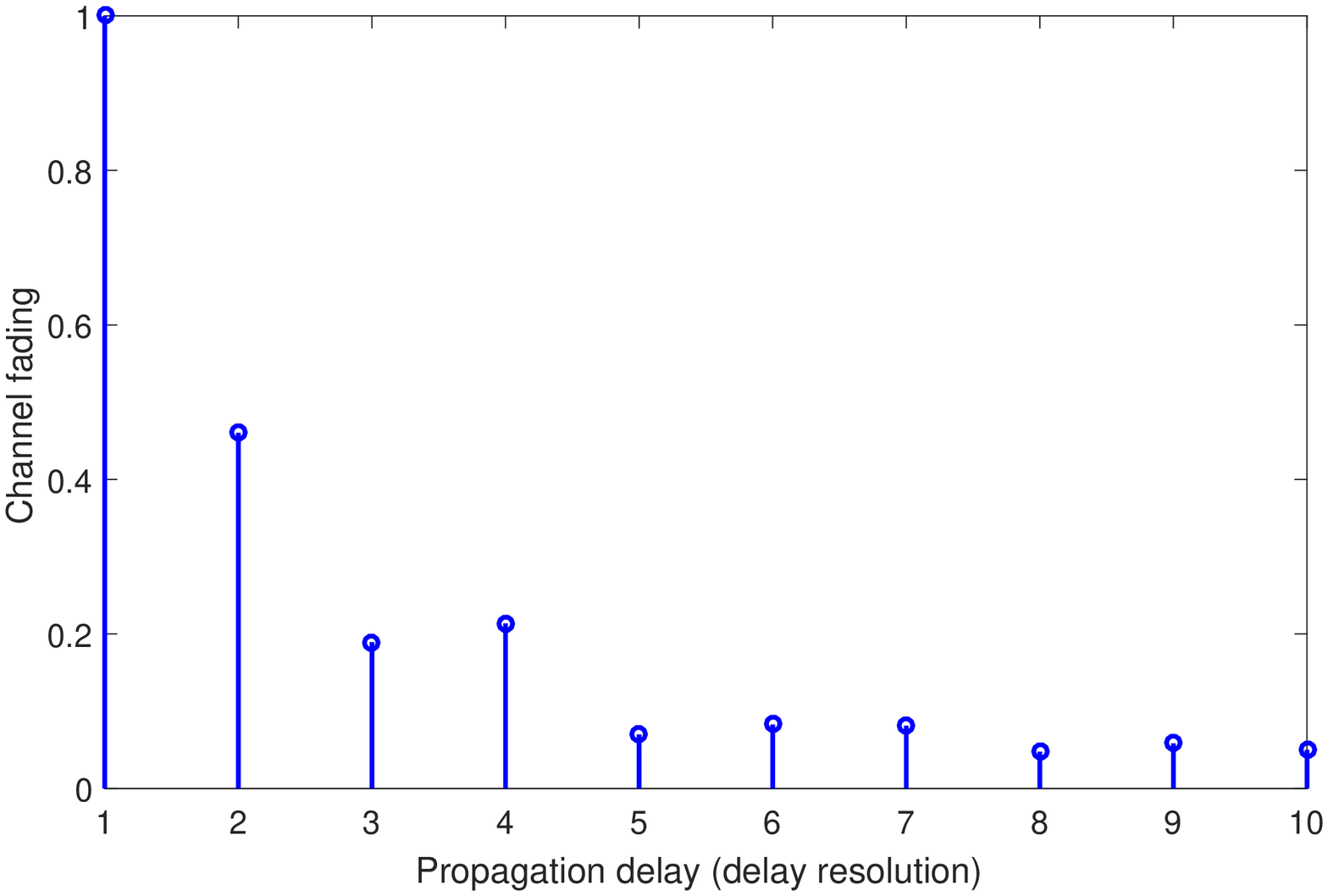}}
\caption{(a) The photo of the test in the second scenario,\ \ \    (b) the normalized estimated channel paramters.}
\label{Scenario2PathTest}
\end{figure*}

\begin{figure}[htb]
\centering{}\includegraphics[scale=0.35]{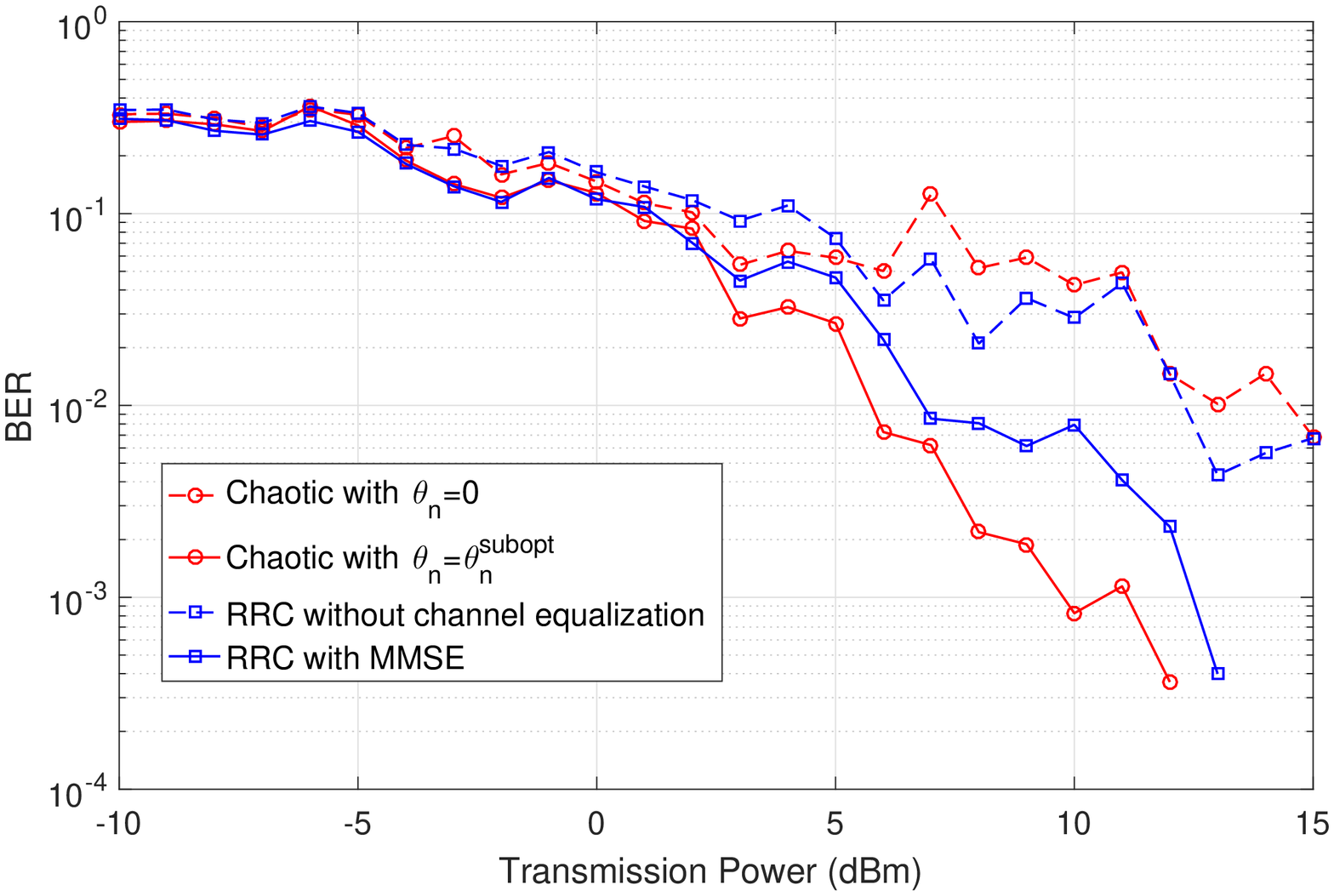}\caption{The experimental BERs of three methods in the second scenario}\label{BERSecondScenario}
\end{figure}

The photo of the test in the first scenario is shown in Fig. \ref{Scenario1PathTest}(a). In this scenario, the distance between TX and RX is about 30 meters and there are some tall buildings in between. The channel parameters are estimated using the LS algorithm, and the normalized estimated channel parameters are shown in Fig. \ref{Scenario1PathTest}(b). We can see that there are three main paths with channel fading 1, 0.31 and 0.17, respectively. Because the delay resolution of our experimental system is 0.125$\mu$s, the delay of three paths are 0$\mu$s, 0.125$\mu$s and 0.25$\mu$s, respectively.

The BERs are tested for four methods: the proposed method using the chaotic baseband waveform and the suboptimal threshold, the proposed method using the chaotic baseband waveform and zero as threshold, the conventional method using RRC without channel equalization, and the conventional method using RRC and MMSE. In WARP V3, the transmission power can be adjusted by parameters TX$_-$RF$_-$Gain and TX$_-$BB$_-$Gain, and the maximum power is 20dBm. We adjust the transmission power to simulate the signal to noise ratio variation. The experimental BER versus transmission power is shown in Fig. \ref{BERFirstScenario}. We can see from Fig. \ref{BERFirstScenario} that the proposed chaotic baseband waveform using zero threshold and the conventional method without channel equalization have the worst BER among the four methods, because the ISIs are strong in this case. However, even with the suboptimal threshold, the performance improvement is significant for the chaotic baseband waveform. For BER=$10^{-3}$, the required transmission power is about 2dB lower than that of the conventional wireless communication system using RRC and MMSE.

The photo of the second test scenario is shown in Fig. \ref{Scenario2PathTest}(a), the distance between TX and RX is about 25 meters, there are some trees, cars and buildings in between, and there is not a line of sight between them. The normalized estimated channel parameters are shown in Fig. \ref{Scenario2PathTest}(b), there are four main paths with channel fading 1, 0.46, 0.19 and 0.21, respectively. The four methods in the first scenario are tested for comparison, and the experimental BERs versus the transmission power are shown in Fig. \ref{BERSecondScenario}. We can see from Fig. \ref{BERSecondScenario} that the chaotic baseband waveform with suboptimal threshold has the best performance, the required transmission power for BER=$10^{-3}$ is about 2.5dB lower than that of the conventional method using RRC and MMSE. The chaotic baseband waveform with zero as threshold and the conventional method without channel equalization have the worst BER, because the ISIs are not relieved in these cases.

Discussion 2. The proposed method needs the past symbols for calculating the suboptimal threshold. However, it's not necessary to use all of the past symbols, because $r(n-m-\tau_l)$ in Eq. (\ref{SubOptimalThreshold}) is close to zero for large enough $n-m-\tau_l$. In the implementation of our experiment, the past $5+\lceil \tau_{L-1} \rceil$ symbols are used to calculate the suboptimal threshold, where $\tau_{L-1}$ is the maximum channel delay. Initially, the training symbols are used to initialize our experimental system.

Discussion 3. Both the proposed method and the conventional method are implemented using the same system configuration showing that our proposed method is compatible with the conventional system, but achieving better BER performance.

Discussion 4. The algorithm complexity of the proposed method is lower than the conventional method with MMSE equalization, because the threshold calculation is simpler as compared to the MMSE algorithm.

\section{Conclusions}
In this paper, a new idea of using chaos as wireless communication baseband waveform is proposed and experimental results validate its effectiveness and superiority. In order to guarantee that the information can be retrieved from the received signal, a continuous-time chaotic waveform is shown to be topologically conjugate to the encoding symbolic dynamic (information). Based on this, a system structure of wireless communication using the chaotic baseband waveform is developed, in which the chaotic shaping filter and the corresponding matched filter are used to implement the encoding and maximize SNR in decoding process at the receiver. Under wireless multipath channel, the method in \cite{Yao2017PRE} is adopted to relieve the ISIs. The simulation comparisons with the conventional wireless communication system show that the proposed method has better performance under both the static and time-varying wireless channel. The experimental results show that, for BER=$10^{-3}$, the required transmiting power of the proposed method is 2dB$\sim$2.5dB lower than that of the conventional method. The merits of the proposed wireless communication system using chaotic baseband waveform include that: i) the encoding method is simpler than the existing impulse control based methods; ii) it is compatible with the commonly-used wireless communication equipment, and can be applied without changing the hardware of the existing digital wireless communication systems; iii) it achieves lower BER than the conventional non-chaotic baseband waveform under time-varying wireless multipath channel. Thus, chaos baseband wireless communication method proves to be a competitive alternative for a conventional wireless communication system to be used in narrowband wireless communication application with high bandwidth efficiency.

\section*{Acknowledgments}
This work was supported by NSFC under Grant No. 61401354, No. 61172070, and No. 61502385; by the Innovative Research Team of Shaanxi Province under Grant No. 2013KCT-04; by Key Basic Research Fund of Shaanxi Province under Grant No. 2016ZDJC0067;  by Natural Science Basic Research Plan in Shaanxi Province of China under Grant No. 2016JQ6015; and by the Foundation of Shaanxi Educational Committee under Grant No.17JS086.

\bibliography{reference}

\end{document}